# Spookytechnology and Society

**CHARLES TAHAN** is a NSF Distinguished International Postdoctoral Research Fellow at the University of Cambridge, Cavendish Laboratory, JJ Thomson Ave, Cambridge, CB3 0HE, United Kingdom.
e-mail: ct320@cam.ac.uk, charlie@tahan.com

**New technologies based on the exploitation of so-called "second order" quantum phenomena – such as quantum entanglement – deserve a public-friendly, rational, and sexy name.** *Spookytechnology* **is that unifying term.**

A smorgasbord of quantum clichés testifies to the influence of quantum physics on technology and society in the 20[th] century, from *Quantum Leap*s to Quantum® corporations. And quantum theory still excites. The new field of quantum information science, as one example, promises a unifying language in many areas of physics and computer science as well as fantastic technologies based on the craziness of the universe. Yet the broader hardware terminology remains poorly considered.

Will the new man-made quantum systems also be called "quantum technology"? Or will "quantum information technology," "quantum coherent technology," "(quantum) entanglement-based technology," or "quantronics" be used instead, as recent grant proposals suggest? We need to avoid the mistakes made in the labeling of nanotechnology, but repeat the successes. By good fortune, the inherent non-intuitiveness of quantum physics has created an opportunity, where the science leads the science fiction. The community can decide for itself, now, how to frame these fields. In the process, we can draw attention to the developments in science, education, and history that are more interesting for science and technology studies than the parallels in nanotech.

Here we introduce "spookytechnology" as a unifying term for the new generation of advanced quantum technologies. From historical and motivational perspectives, this name has greater value than the many variations of quantum this and quantum that presently used. At the cocktail party level, *spookytechnology is technology based on the spooky properties of quantum physics*. More technically,

> *spookytechnology* encompasses all functional devices, systems, and materials whose utility relies in whole or in part on higher order quantum properties of matter and energy that have no counterpart in the classical world. These purely quantum traits may include superposition, entanglement, decoherence (along with the quantum aspects of measurement and error correction) or new behavior that emerges in engineered quantum many-body systems.

The term "spookytechnology" has a strong historical foundation: Einstein himself. Albert Einstein, despite being an excellent quantum mechanic, hated certain aspects of quantum theory. He called entanglement "spooky action at a distance." Together with Rosen and Podolsky, his critical but brilliant insights led indirectly to the inception of the spookytechnologies nearly fifty years later. "Spookytechnology" also benefits from being simple, succinct, and specific (as we have defined it) yet broad enough to contain more than just quantum information technology. The term is also inherently interesting, one aspect where "nanotechnology" is a great role model. "Nanotechnology" has always been ambiguous but popular.

Consider the opportunity at the intersection of society and spookytechnology. Nanotechnology is currently of wider interest. For one, spookytechnology poses no direct environmental or toxicological risks. For another, spookytechnology remains in the science inception phase, although commercial products are beginning to appear. Like nanotechnology, necessity has created a fertile, interdisciplinary environment, and experimental progress has made devices realizable only recently. However, the science behind spookytechnology represents a paradigm shift far deeper than "nanoscience," which largely consists of a loose confederation of distinct fields.

Spookytechnology is guided by a rigorous theoretical foundation and specific technical proposals. An increasingly adopted, evolving language founded on quantum optics and information theory allows dialogue between distinct subfields within physics and across to computer science, math, and engineering. This influences education and textbook design. A new conceptual framework for understanding and unifying disparate physical systems (including perhaps the universe itself) is appearing. Why is spookytechnology so interesting?

On the one hand we're talking about fundamental physics – real secret-of-the-universe type stuff – and on the other hand we have the promise of revolutionary technological innovation.

Of course, a quantum technological revolution has happened before. By the late 1920s the quantum theory initiated by Planck became convincing and useful. An understanding of the dual wave and particle-like behavior of nature at the microscopic level led to a revolution of quantum-designed technologies, from the photocopier to the atomic bomb. The description of electrons as waves traveling through a semiconductor crystal or the quantization of light into particles dubbed photons are just two examples. The last 20 years have seen a similar consensus in the theory of quantum information science and quantum coherent or entanglement-based technology. Dowling and Milburn convincingly argue that this constitutes the beginning of a "second quantum revolution".

The spookytechnologies make use of mostly sidestepped quantum phenomena that emerge from the formal theory but still perplex us. These surround the nature of the universe itself, how it is interconnected, and how the quantum world becomes the classical world. Measurement is one example. In contrast to a classical object (like watching an apple fall), a quantum object becomes fundamentally altered after observation. When exactly does the universe stop acting quantum ("decoherence") and start behaving classically? Through exquisite control can we make a system stay "quantum coherent" far into classical territory? A dramatic natural example is superconductivity, a quantum many-body phenomenon where long-range phase coherence allows practically zero electric current resistance at temperatures well above absolute zero.

The king of spooky quantumness is quantum entanglement. As physicists define it, "entanglement" refers to a peculiar quality of the known universe whereby quantum objects in certain situations cannot be described separately, even though they may be separated in space. In other words, if an atom in London is entangled with an atom in Tokyo, they are still one quantum system, which can only be understood together. Entanglement is a multi-object generalization of quantum superposition, another astounding quantum trait, where a particle can exist in two states at once (here *and* there, up *and* down). Both properties have been experimentally confirmed.

Applied to information theory, quantum superposition generalizes a classical digital bit (on or off, 0 or 1) to a *qu*b*it (*any* linear combination of 0 and 1). Superposition and entanglement provide a more general basis for computing: computation based on physical laws, not human abstractions. Intrinsically more efficient quantum algorithms have been proposed and demonstrated on these computing systems. Two discoveries made in the mid 1990s – an exponentially faster quantum algorithm for code breaking and the possibility of quantum error correction – began a serious, government-based investment campaign in quantum information technology. Yes, "the spooks" (NSA, etc.) fund a lot of spookytechnology. The control and ability needed to construct a large quantum computer device is the extreme example of quantum technology engineering based on the spooky properties of the quantum world.

Spookytechnology covers more than quantum computers. Comparatively simple cryptographic systems based on quantum entanglement have already achieved practical deployment. Beyond information technology, specialized hardware devices or systems based on quantum coherent control or entanglement are also in development. Consider the case of what some call "quantum imaging." An example where the super-saturated "quantum" becomes a detriment, "quantum imaging" can refer to new ideas for entanglement-based sub-wavelength imaging and lithography. Yet magnetic resonance imaging (MRI) departments across the world have been called "quantum imaging" groups long before the advent of entanglement-based design ideas. MRI has always been quantum, involving the manipulation of spins in the body, which are purely quantum aspects of atomic nuclei or electrons. These quantum possibilities weren't dreamed of when nuclear magnetic resonance was conceived. "Spooky imaging" may be more appropriate.

"Spookytechnology" as a useful term is open to critique. New words are hard to justify until they've already happened. But there is value in a term that reframes and separates the new quantum technologies from the old ones. Another standard might be chosen; "quantum coherent technology" is probably the most inclusive alternative. Yet "spookytechnology" is alluring and points to what is unique about these technologies. As we have shown, the alternatives are either too narrow, overused to oblivion, or too long and complicated. We don't necessarily want a word that creates crazy speculation or adds to the ignorance of the general population. But quantum already does this. In general, more interest is better than less. If spookytechnology better invites the question: "What is that?," than it has more educational value.

Nor do we want to incite a prefix-fest as in nano-*everything*. "Spooky," being defined more specifically, has fewer tendencies towards this than "nano," which alludes to an entire length scale. Terms like "spookynet" or "spookytronics" *may* make sense, but selectively. Defined outside of science and later co-opted, "nanotechnology" became a broad term for everything from advanced materials, to mesoscopics, to small robots, to genetic engineering. Since there is no authoritative initial definition to look up, classification (for commercial, toxicological, or patent law purposes) becomes difficult. But nanotechnology is also a positive example of the involvement of society in a new field; of the value it can have to science education, to science and technology studies, and as an investment focal point. Spookytechnology, defined formally and narrowly, can guide us through the positives of such a transformation and minimize the negatives.

We do not try to usurp the great and long previous work glorifying and explaining quantum computing. Instead, we propose a unifying word for a broader set of technologies that will include quantum computers. We attempt to lay a framework for work in science and technology studies regarding this new technological domain, from a perspective relevant to the nanotechnology and society industry. One could argue that all technologies are spooky when they first appear – from Newtonian gravity's "action at a distance" to the mysterious waves explored by Tesla and Marconi (microwave and radio). But quantum physics will never become intuitive, even if it becomes commonplace. If we put 1995 as the "this can be done" moment for spookytechnology, that would put a 2010-15 time frame to mainstream industry penetration. The scope and depth of these historical and scientific developments should not go unrecognized. Spookytechnology will find its place in the increasingly dense line of major technological revolutions began with the industrial revolution in the eighteenth century: quantum, info, bio, nano, spooky.

## SELECTED FURTHER READING (NON-TECHNICAL)

## ACKNOWLEDGEMENTS

The author is supported by a USA National Science Foundation Math and Physical Sciences Distinguished International Postdoctoral Research Fellowship (Award No. DMR-0502047). Thanks to Greta Zenner, Mark Friesen, and Clark Miller for critical reading.

*Be not afeared.*
*The world is full of enormous opportunities,*
*Changes and challenges*
*That give delight and hurt not.*
                                        William Shakespeare